# Fog Computing based SDI Framework for Mineral Resources Information Infrastructure Management in India

Rabindra K. Barik
KIIT University,
Bhubaneswar, India
rabindra.mnnit@gmail.com

Rakesh K. Lenka
IIIT-Bhubaneswar,
Bhubaneswar, India
rakeshkumar@iiit-bh.ac.in

N.V.R. Simha
IIIT-Bhubaneswar,
Bhubaneswar, India
a116012@iiit-bh.ac.in

Harishchandra Dubey
University of Texas
Dallas, USA
harishchandra.dubey@utdallas.edu

Kunal Mankodiya
University of Rhode Island,
USA
kunalm@uri.edu

*Abstract*— **Spatial Data Infrastructure (SDI) is an important concept for sharing spatial data across the web. With cumulative techniques with spatial cloud computing and fog computing, SDI has the greater potential and has been emerged as a tool for processing, analysis and transmission of spatial data. The Fog computing is a paradigm where Fog devices help to increase throughput and reduce latency at the edge of the client with respect to cloud computing environment. This paper proposed and developed a fog computing based SDI framework for mining analytics from spatial big data for mineral resources management in India. We built a prototype using Raspberry Pi, an embedded microprocessor. We validated by taking suitable case study of mineral resources management in India by doing preliminary analysis including overlay analysis. Results showed that fog computing hold a great promise for analysis of spatial data. We used open source GIS i.e. QGIS and QIS plugin for reducing the transmission to cloud from the fog node.**

## I. Introduction

Spatial Data Infrastructure (SDI) has facilitated by sharing and exchanging of spatial data holding by various stake-holders. It has initiated to create an environment which enables a wide variety of users to retrieve, access and disseminate spatial and related non-spatial data in secure way[1]. SDI has the capability for storage and decision making of spatial data, bringing data and maps to a common scale as per the user needs, querying, superimposing and analyzing the data and designing/ presenting final reports/ maps to the administrators and planners [2]. The utility of SDI for planning of environmental monitoring, natural resource management, health care, land use planning and urban planning, watershed management, marine or coastal managements [3][1], land information and decision making have become widely popular and are being used in different areas for a wide range of applications. SDI has became an emerging area in which it has the ability to integrate and analyze heterogeneous thematic layers along with their attribute information to create and visualize alternative planning scenarios for planners and decision makers. The user friendliness of SDI is a feature that has made SDI a preferred platform for planning in global, regional, national and local level coupled with various analyses and modeling functionalities. SDI integrates common spatial database operations such as query formation, statistical computations and overlay analysis with unique visualization and geographical functionalities[3][4]. These characteristics distinguish SDI from other spatial decision support systems and make it valuable to a wide range of public and private enterprises for explaining events, predicting outcomes and designing strategies. By integrating SDI with cloud computing technology and fog computing , it has been merged to perform a value added services that give rise to spatial cloud computing [9] [13] [18]. Spatial data have rich information about temporal as well as spatial distributions[14]. In traditional setup, we send the data to the cloud server where these are going for further processing and analysis.

## II. Proposed Framework

In the present paper, it has proposed the advance architecture of spatial fog computing in mineral resources management sector. The mineral resources related spatial data has been processed at the edge using Fog computing device. The present paper made the following contributions to mineral resources management:

- mrFog framework is proposed for improved throughput and reduced latency for analysis and transmission of spatial data in mineral resources management sector
- Raspberry Pi was employed as the fog computing device.
- Spatial data analysis scheme and overlay analysis in thin clients environment was performed using mrFog Framework.
- It has performed a case study by doing overlay analysis of mineral resources management in India.

## III. Related work

### *A. SDI Framework*

In early 80s and late 70s, many national surveying and mapping agencies has planned for coordination of surveying and mapping activities. They felt the need to initiate strategy for providing greater access to develop Geographical Information tools and efforts to build spatial information system [6], [7], [24], [25]. In this process, focus has been shifted on the conceptualization and development of SDI) The term SDI has coined in 1993 by the US National Research Council to describe the provision of standardized geographical information access. SDI constitutes a set of relationships and partnerships which enable the spatial data sharing, updating and integration [5].One of the core components of SDI framework can be defined as data people, access networking, policy and standards. These can be grouped into different categories based on the nature of their interactions within the SDI framework [8]. The people and data can be taken as one category while the second category can be the access network, standard and polices as the main technological components. The nature of second category is dynamic due to the rapid development of technology [10]. This has been suggested that an integrated SDI cannot be thought as having spatial data, end-users alone and value-added services but instead involves other important issues regarding scalability, interoperability, network, security and policies also. That has been reflected the dynamic nature of SDI framework concept that is depicted in Figure 1.

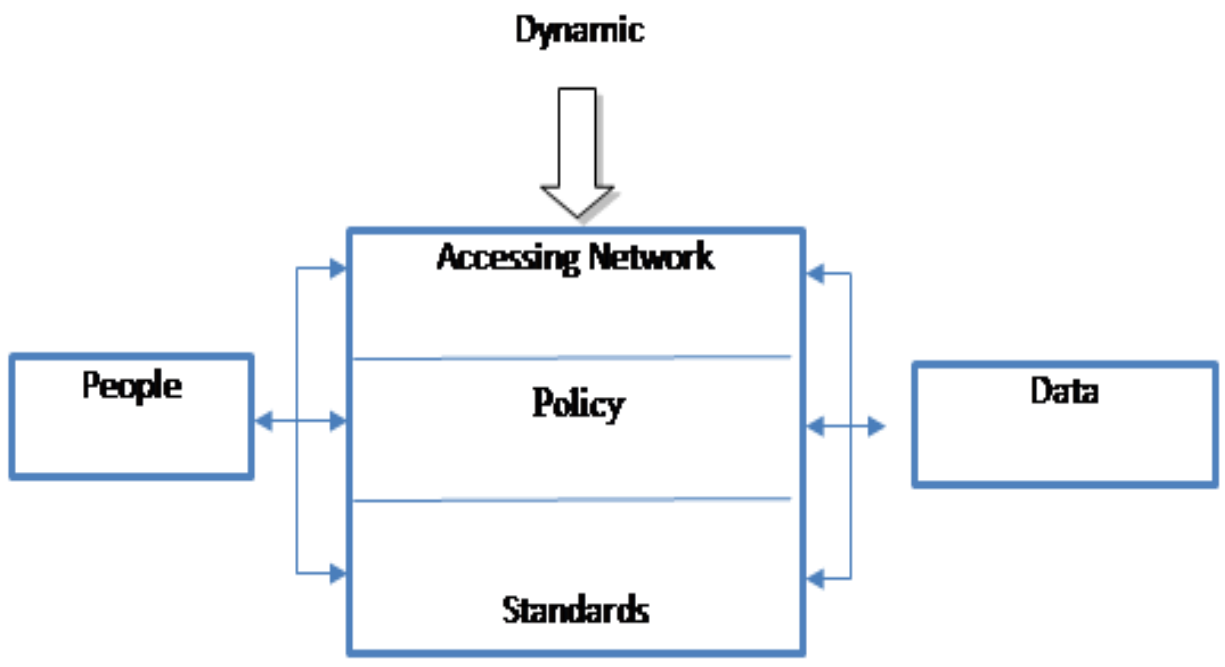

*Fig.1* The Dynamic Nature of SDI Framework [1]

From the dynamic nature of SDI framework, it has been defined as to share data across the global, regional, national, state and local levels. By advancement of technology of Service Oriented Architecture (SOA) and cloud computing framework, there is a huge development in the spatial data sharing and processing mechanism in SDI framework [11][12]. The cloud computing and fog computing technology have been used for implementing in SDI framework [19][20] [21][18] for better management of spatial data over web, that give rise to the concept of mineral resources management in India which has been discuss in the next section.

### *B. Need of SDI for mineral resources management in India*

Mineral Resource exploration and management are an important activity that can contribute considerably to the economy of India as well as foreign investments. This mining sector has been facing an increased competition on account of globalization and the level of technologies which has been employed. Therefore, it is necessitating a new beginning for meeting the innovative challenges of better extraction and management of mineral resources. Mineral resources are precious and non-renewable and limited, which has been required a unique management policy and standardization. The scientific managements and proper standardization policy not only provide the efficient use and exploitation of mineral resources but also relates to social, economical and sustainable development. That leads to make proper coordinated efforts to support greater investment in exploration and management of mineral re-sources. By integrate cloud and fog computing technologies with Remote Sensing, GIS and GPS technology to increase efficiency and productivity of this mineral sector. That needs to establish a well organized mineral resources information Infrastructure management in India with the help of fog computing and spatial technology. Thus, Fog computing based SDI framework for mineral resources information infrastructure from the perspective of user is one of the urgent needs to the society. In present era of high end information scenario, new technologies and tools have emerged to store, collect, analyses and retrieve various types of information related to mineral resources de-posits. These unique challenges cause barriers in extensively information sharing of mineral resources data and restrain the effectiveness in responding and understanding to proper management. To overcome these challenges in mining sec-tor, sharing and mapping of mineral resources information in an interoperable and secure framework under GIS and fog computing environment [18]. So, next section describes the fog computing approaches in different scenario.

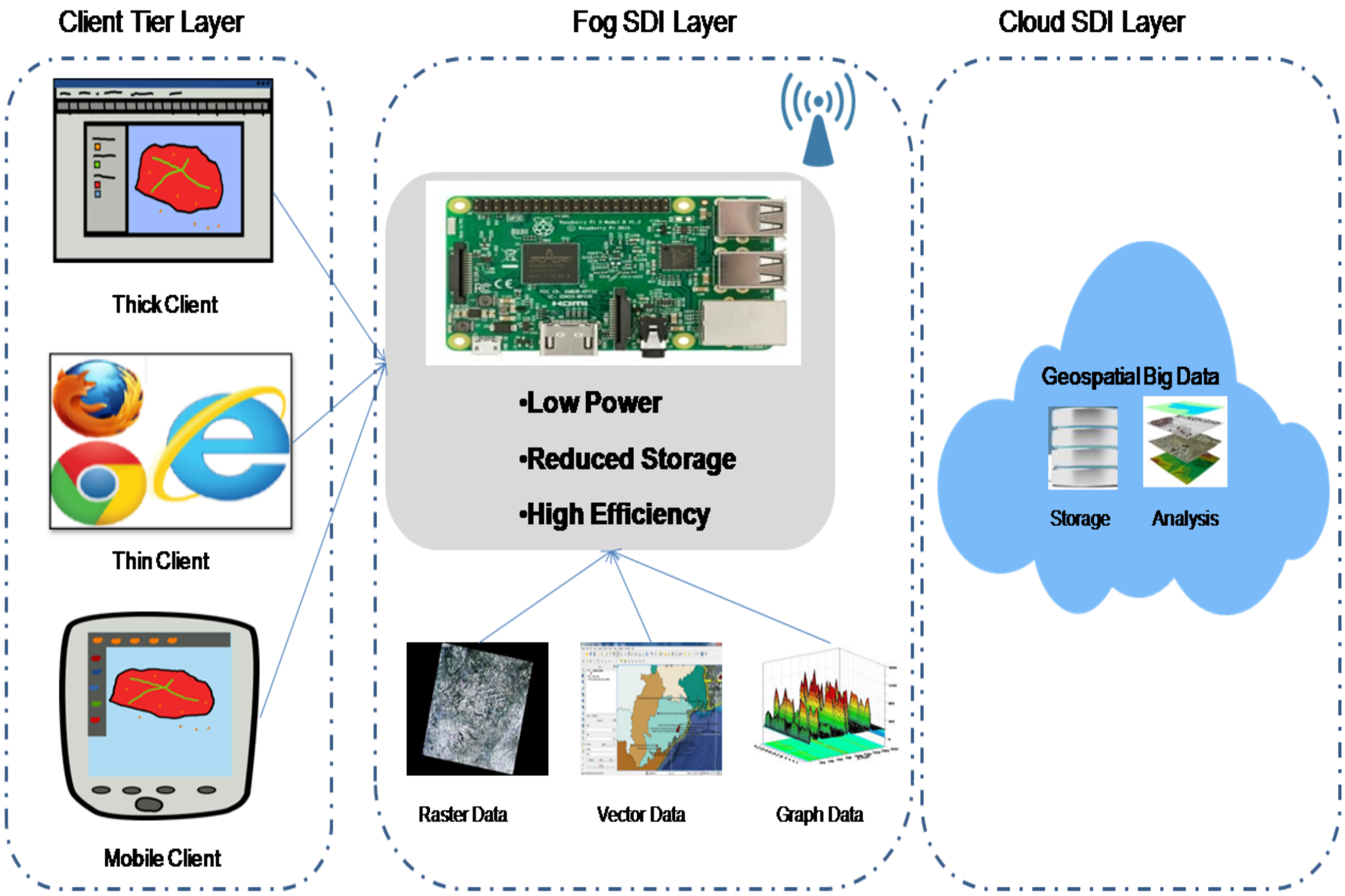

*Fig.2.* mrFog: Conceptual diagram of the proposed FogSDI framework for power-efficient, low latency and high throughput analysis of the spatial data.

## IV. Proposed Model

### *A. Fog Computing*

Cloud computing provides shared data analysis and computer processing, in other terms Cloud is a hub of computing resources such as high power computing servers, computer networks, storage and services. The availability of high-capacity networks, low-cost computers and storage devices makes cloud a highly demanded service for the users seeking for high computing power. Cloud can interact with the fog devices. From the last couple of years, Fog Computing has been provided the low-power gateway which can increase throughput and has been reduced the latency near the edge of the various spatial data processing at client layer [15][16][23][26]. It has also found that the fog computing reduces the storage needed for spatial big data in the cloud computing environment. In addition, the reduction of storage space in the required transmission power results in overall improvement and efficiency of its uses. Fog computing has been successfully implemented in many sector like health-care and smart cities [17][18][22]. This paper concentrates on the side of fog computing, which allows users a higher computing power at the instrument end. . Reliance on fog will help cut the costs associated with the Cloud to an extent in the area of mineral resources management. Thus, by integrating of fog computing with SDI technology, we proposed a framework which is discussed in next section.

### *B. mrFog Framework*

This section describes various components of the proposed Fog SDI Framework i.e. mrFog and discusses the methods implemented in it. It has been employed Raspberry Pi as Fog computing device in proposed mrFog framework. Raspberry Pi connects to WIFI and has been used Ubi-Linux operating system for running compression utilities. Figure2 shows the proposed mrFog Framework. The fog device acts as a gateway between thick, thin and mobile clients and cloud layer. The proposed Fog SDI Framework has three layers as client tier layer, cloud SDI layer and Fog SDI layer. In client tier, the categories of users have been further divided into thick client, thin client and mobile client environment.

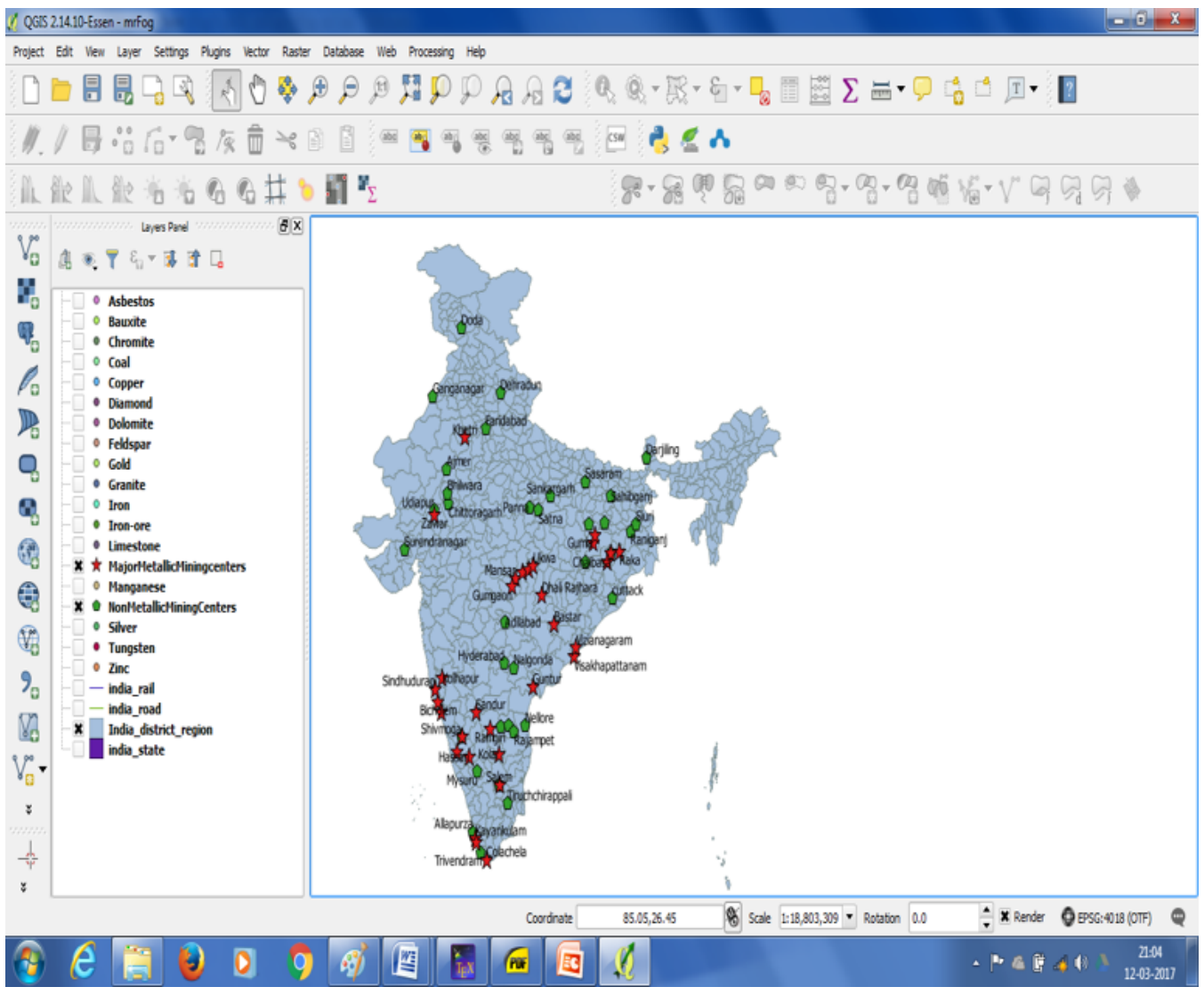

*Fig.3. Overlay operation on thick client environment in Quantum GIS Environment.*

Processing of spatial data can be possible within these three environments. Spatial Cloud layer is mainly focused on overall storage and analysis of spatial data. The Fog layer works as middle tier between client tier layer and spatial cloud layer. It has been experimentally validated that the fog layer is characterized by low power consumption, reduced storage requirement and overlay analysis capabilities.

## V. RESULTS & DISCUSSIONS

In this section, data analysis particularly overlay analysis is performed for mineral resources mapping of India. Over-lay Analysis has been done with the help of QGIS Plugin Quantum GIS open source GIS environment. It has been superimposed various spatial data in a common platform for better analysis of mineral spatial data. It also gives the platform to analyze the pattern and trends of occurrence of minerals from the different regions of India. Figure 3 shows the snapshot of the overlay analysis of mineral information mapping of India in Quantum GIS environment

## VI. CONCLUSION

In this paper, it has been developed mrFog framework that employed Fog gateway in the middle of Client tier and Cloud SDI tier. Raspberry Pi processor has been used as the fog computer. The Fog gateway reduces the transmission power, storage space requirements and increased throughput to overall efficiency of mineral resources management using mrFog framework. The mrFog framework introduces edge intelligence in mineral resources spatial cloud environment. In future, it has to compare and contrast with the cloud SDI architecture with addition of more intelligent processing at the Fog layer in mobile client environments.

## REFERENCES

[1] Georis-Creuseveau J, Claramunt C, Gourmelon F. A modelling framework for the study of Spatial Data Infrastructures applied to coastal management and planning. International Journal of Geographical Information Science. 2017 Jan 2;31(1):122-38.

[2] Smith J, Mackaness W, Kealy A, Williamson I. Spatial data infrastructure requirements for mobile location based journey planning. Transactions in GIS. 2004 Jan 1;8(1):23-2.

[3] Giuliani G, Lacroix P, Guigoz Y, Roncella R, Bigagli L, Santoro M, Mazzetti P, Nativi S, Ray N, Lehmann A. Bringing GEOSS Services into Practice: A Capacity Building Resource on Spatial Data Infrastructures (SDI). Transactions in GIS. 2016 May 1.

[4] Coleman DJ, Rajabifard A, Kolodziej KW. Expanding the SDI environment: comparing current spatial data infrastructure with emerging indoor location-based services. International Journal of Digital Earth. 2016 Jun 2;9(6):629-47.

[5] Yue P, Guo X, Zhang M, Jiang L, Zhai X. Linked Data and SDI: The case on Web geoprocessing workflows. ISPRS Journal of Photogrammetry and Remote Sensing. 2016 Apr 30;114:245-57.

[6] Pun-Cheng LS, Lai WW, Chang RK. Exploring utility system SDI–Managerial and technical perspectives. Tunnelling and Underground Space Technology. 2016 Apr 30;54:13-9.

[7] Koswatte S, McDougall K, Liu X. SDI and crowdsourced spatial information management automation for disaster management. Survey Review. 2015 Sep 1;47(344):307-15.

[8] Salajegheh J, Hakimpour F, Esmaeily A. Developing a web-based system by integrating VGI and SDI for real estate management and marketing. The International Archives of Photogrammetry, Remote Sensing and Spatial Information Sciences. 2014 Jan 1;40(2):231.

[9] Schäffer B, Baranski B, Foerster T. Towards spatial data infrastructures in the clouds. Geospatial thinking. 2010:399-418.

[10] Mwange C, Mulaku GC, Siriba DN. Reviewing the status of national spatial data infrastructures in Africa.
Survey Review. 2016 Dec 1:1-0.

[11] Barik RK, Samaddar AB. Service Oriented Architecture Based SDIModel for Education Sector in India. InProceedings of the International Conference on Frontiers of Intelligent Computing: Theory and Applications (FICTA) 2013 2014 (pp. 555-562). Springer International Publishing.
.

[12] Barik RK, Samaddar AB. Service oriented architecture based sdi model for mineral resources management in india. Universal Journal of Geoscience. 2014 Jan;2(1):1-6.

[13] Yang C, Huang Q, Li Z, Liu K, Hu F. Big Data and cloud computing: innovation opportunities and challenges. International Journal of Digital Earth. 2017 Jan 2;10(1):13-53.

[14] Lee JG, Kang M. Geospatial big data: challenges and opportunities. Big Data Research. 2015 Jun 30;2(2):74-81.

[15] Dubey H, Goldberg JC, Abtahi M, Mahler L, Mankodiya K. EchoWear: smartwatch technology for voice and speech treatments of patients with Parkinson's disease. InProceedings of the conference on Wireless Health 2015 Oct 14 (p. 15). ACM.

[16] Dubey H, Monteiro A, Mahler L, Akbar U, Sun Y, Yang Q, Mankodiya K. FogCare: fog-assisted internet of things for smart telemedicine. Future Gen Comput Syst. 2016.

[17] Monteiro A, Dubey H, Mahler L, Yang Q, Mankodiya K. Fit: A fog computing device for speech tele-treatments. InSmart Computing (SMARTCOMP), 2016 IEEE International Conference on 2016 May 18 (pp. 1-3). IEEE.

[18] Barik RK, Dubey H, Samaddar AB, Gupta RD, Ray PK. FogGIS: Fog Computing for Geospatial Big Data Analytics. arXiv preprint arXiv:1701.02601. 2016.

[19] Evangelidis K, Ntouros K, Makridis S, Papatheodorou C. Geospatial services in the Cloud. Computers & Geosciences. 2014 Feb 28;63:116-22.

[20] Kharouf, R.A.A., Alzoubaidi, A.R. and Jweihan, M., An integrated architectural framework for geoprocessing in cloud environment. Spatial Information Research, pp.1-9.

[21] Lenka RK, Barik RK, Gupta N, Ali SM, Rath A, Dubey H. Comparative Analysis of SpatialHadoop and GeoSpark for Geospatial Big Data Analytics. arXiv preprint arXiv:1612.07433. 2016 Dec.

[22] Constant N, Borthakur D, Abtahi M, Dubey H, Mankodiya K. Fog-assisted wiot: A smart fog gateway for end-to-end analytics in wearable internet of things. arXiv preprint arXiv:1701.08680. 2017 .

[23] Dubey H, Yang J, Constant N, Amiri AM, Yang Q, Makodiya K. Fog data: Enhancing telehealth big data through fog computing. InProceedings of the ASE BigData & SocialInformatics 2015 2015 Oct 7 (p. 14). ACM.

[24] Barik RK, Das PK, Lenka RK. Development and implementation of SOA based SDI model for tourism information infrastructure management web services. InCloud System and Big Data Engineering (Confluence), 2016 6th International Conference 2016 Jan 14 (pp. 748-753). IEEE.

[25] Barik RK, Samaddar AB, Gupta RD. Investigations into the Efficacy of Open Source GIS Software, MapWorld Forum, 2009.

[26] Ketel M. Fog-Cloud Services for IoT. InProceedings of the SouthEast Conference 2017 Apr 13 (pp. 262-264). ACM.